\documentclass[aps,prl,10pt,twocolumn,floatfix,superscriptaddress]{revtex4-2}

\usepackage{times}
\usepackage{graphicx}
\usepackage{dcolumn}
\usepackage{bm}
\usepackage{xcolor}
\usepackage{mathtools}
\usepackage{bbold}
\usepackage{changes}

\usepackage{hyperref}
\hypersetup{
colorlinks=true,
linkcolor=blue,           
citecolor=blue,           
filecolor=magenta,        
urlcolor=cyan
}

\renewcommand{\i}{\mathrm{i}}

\DeclarePairedDelimiter\ket{\lvert}{\rangle}
\DeclarePairedDelimiterX\braket[2]{\langle}{\rangle}{#1 \delimsize\vert #2}
\DeclarePairedDelimiterX\ketbra[2]{\lvert}{\rvert}{#1 \rangle\hspace{-.25em}\langle #2}

\begin{document}

\title{Correspondence between non-Hermitian topology and directional amplification in the presence of disorder}

\author{Clara C.~Wanjura}
\affiliation{Cavendish Laboratory, University of Cambridge, Cambridge CB3 0HE, United Kingdom}

\author{Matteo Brunelli}
\affiliation{Cavendish Laboratory, University of Cambridge, Cambridge CB3 0HE, United Kingdom}

\author{Andreas Nunnenkamp}
\affiliation{
School of Physics and Astronomy and Centre for the Mathematics and Theoretical Physics of Quantum Non-Equilibrium Systems, University of Nottingham, Nottingham, NG7 2RD, United Kingdom}
\affiliation{Cavendish Laboratory, University of Cambridge, Cambridge CB3 0HE, United Kingdom}

\date{\today}

\begin{abstract}
In order for non-Hermitian (NH) topological effects to be relevant for practical applications, it is necessary to study disordered systems. In the absence of disorder, certain driven-dissipative cavity arrays with engineered non-local dissipation display directional amplification when associated with a non-trivial winding number of the NH dynamic matrix. In this work, we show analytically that the correspondence between NH topology and directional amplification holds even in the presence of disorder.  When a system with non-trivial topology is tuned close to the exceptional point, perfect non-reciprocity (quantified by a vanishing reverse gain) is preserved for arbitrarily strong on-site disorder. For bounded disorder, we derive simple bounds for the probability distribution of the scattering matrix elements. These bounds show that the essential features associated with non-trivial NH topology, namely that the end-to-end forward (reverse) gain grows (is suppressed) exponentially with system size, are preserved in disordered systems. NH topology in cavity arrays is robust and can thus be exploited for practical applications.
\end{abstract}

\keywords{topology, directional amplification, non-reciprocity, reservoir engineering, disorder, topological protection, robustness}

\maketitle

\textit{Introduction.---}%
One of the central properties of topological transport is that it is carried by chiral edge states and protected against back-scattering and disorder~\cite{Hasan2010, Bansil2016}.
In the case of photons, this has led to much interest in topological waveguides, amplifiers, and lasers, and given birth to the field of topological photonics~\cite{Ozawa2019}.
Photonic systems typically experience gain and loss, which are naturally accounted for by effective non-Hermitian (NH) Hamiltonians.
Accordingly, describing their topological features requires going beyond the standard (Hermitian) characterization, as witnessed by the surge of activity in the newborn field of NH topology~\cite{Bergholtz2019}.
For a class of driven-dissipative cavity arrays, featuring engineered non-local dissipation, we have recently shown that the occurrence of directional end-to-end amplification is associated with a non-trivial topological winding number of the (NH) dynamic matrix; directional amplification can thus be understood as NH topological amplification~\cite{Wanjura2020}.

A complete characterization of topological transport cannot prescind from discussing the role of disorder. While for Hermitian topological systems this is a common practice, the role of disorder in NH topological systems---and the potential robustness against it---is much less explored. Previous studies on the effects of disorder in NH topological systems are mostly numerical and have focused on spectral properties~\cite{Alvarez2018, Gong2018, Yuce2019, Rivero2020, Ashida2020}, topological invariants~\cite{Gong2018, Claes2020}, and exceptional points~\cite{Yuce2019, Rivero2020}. 
\begin{figure}[t]
\centering
\includegraphics[width=\linewidth]{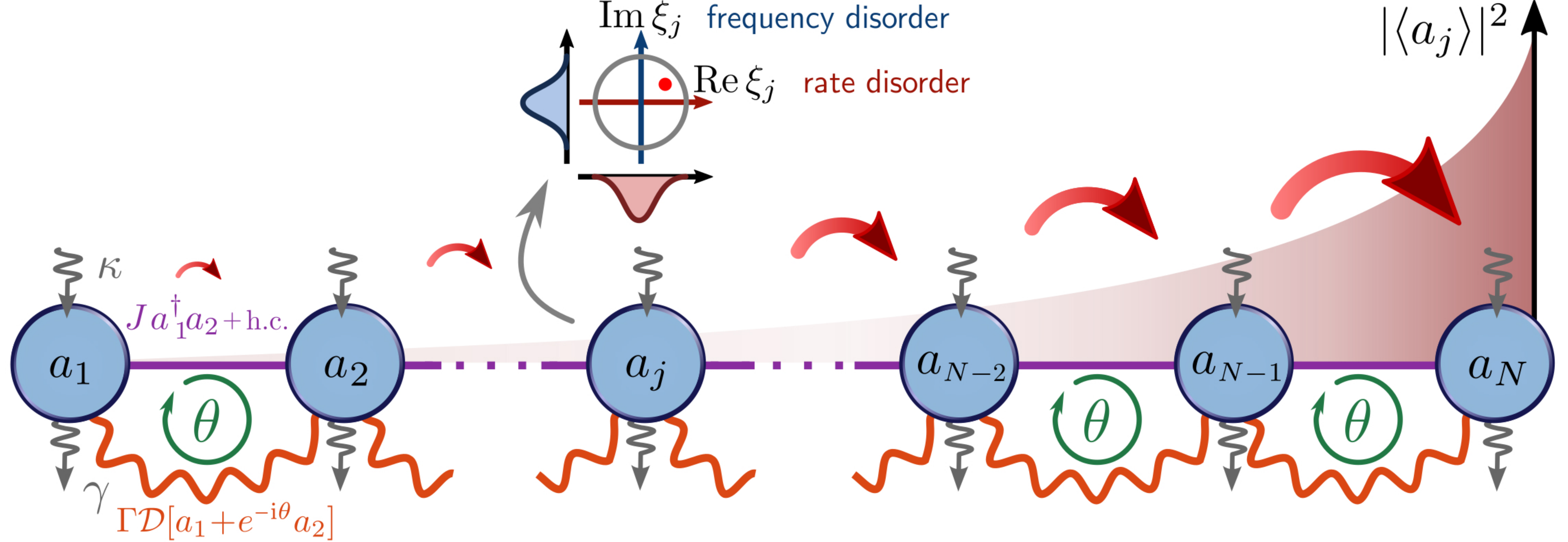}
\caption{\textbf{Driven-dissipative cavity array with on-site disorder.}
Adjacent sites are coupled coherently by beamsplitter interactions (purple straight lines) and dissipatively by non-local dissipators (wiggly orange lines). These two together establish a gauge-invariant phase $\theta$, which controls non-reciprocity in the array. Each cavity is subject to gain (incoming arrows) and loss (outgoing arrows). Gain and loss, as well as cavity frequencies, are subject to disorder $\xi_j$ at each site, i.e., their values are randomly distributed (see inset). Without disorder, topologically non-trivial regimes correspond to directional amplification (red shaded area). We show that this extends to disordered systems.}
\label{fig:physicalSystem}
\end{figure}

In this work we study NH topological amplification in driven-dissipative cavity arrays subject to on-site disorder. Our analysis goes beyond the study of the complex spectrum and investigates the effects of disorder on the  scattering matrix, which characterizes the transport properties of the system. When a system with non-trivial topology is tuned to (or close to) the exceptional point (EP), we show that perfect non-reciprocity---quantified by a vanishing reverse gain---is preserved for any kind of disorder; the disorder can be arbitrarily strong and have arbitrary distribution. For the general case, we show that topological amplification is robust against disorder by placing bounds on the probability distribution of the scattering matrix elements, in particular, the end-to-end gain. The bounds have a simple analytic expression and rely on the only assumption that the disorder has compact support. 
The bounds show that the essential features associated with non-trivial NH topology, namely that the end-to-end forward (reverse) gain grows exponentially (is exponentially suppressed) with system size, extend to disordered systems. This holds true for any system size, even in the thermodynamic limit. These scalings prove the robustness of NH topological amplification. Taken together, our results extend the correspondence between non-trivial NH topology and directional amplification~\cite{Wanjura2020} to disordered systems.

Our findings apply to several candidate platforms for the exploration of NH topological physics, where in practice disorder can often not be neglected. These include optomechanical systems~\cite{Aspelmeyer2014, Verhagen2017, Mercier2019}, superconducting circuits~\cite{Abdo2013, Bergeal2010}, and topolectric circuits~\cite{Lee2018, Kotwal2019}. Our results are especially relevant for the design of directional multimode amplifiers and for sensing applications~\cite{Budich2020, McDonald2018}. Our analysis may also provide new insight into the role of disorder in NH systems such as random lasers~\cite{Wiersma2008,Schonhuber2016}, disordered cavities~\cite{Schomerus2009} and random photonics~\cite{Wiersma2013}.

\textit{Model.---}%
As shown in Fig.~\ref{fig:physicalSystem}, we consider $N$ bosonic (cavity) modes, each subject to photon decay with rate $\gamma$ and incoherent pumping with rate $\kappa$.
Coherent beam splitter interactions (with $\hbar=1$) $\mathcal{H} = \sum_j J a_j^\dagger a_{j+1} + \mathrm{h.c.}$ are combined with their dissipative counterpart, described by the non-local dissipator $\mathcal{D}[z_j]\equiv z_j\rho z_j^\dagger-\frac{1}{2}\{z_j^\dagger z_j,\rho\}$, with $z_j\equiv a_j + e^{-\i\theta} a_{j+1}$ and rate $\Gamma$, so that the evolution of the system density operator $\rho$ is determined by the master equation
\begin{align}\label{MasterEq}
\dot \rho & = -\i [\mathcal{H},\rho] + \sum_j \left(\Gamma \mathcal{D}[z_j]\rho+\gamma \mathcal{D}[a_j]\rho + \kappa \mathcal{D}[a_j^\dagger]\rho\right).
\end{align}
Balancing coherent and dissipative interactions $(\vert J\vert=\Gamma/2)$ and selecting a specific value of the phase $(\theta=\frac{\pi}{2},\frac{3\pi}{2})$, gives rise to perfect unidirectional transmission~\cite{Metelmann2015, Metelmann2017}, for which standard theory of cascaded quantum systems is recovered~\cite{Carmichael1993,Gardiner1993}. Deviations from this condition still yield non-reciprocal dynamics apart from $\theta=0,\pi$ when the dynamics are fully reciprocal. The presence of a gain mechanism (incoherent pump), acting on top of the non-reciprocal dynamics, enables directional amplification.
Eq.~\eqref{MasterEq} gives the following equations of motion for the mean cavity amplitudes $\langle a_j\rangle$~\cite{Wanjura2020}
\begin{align}
\langle\dot a_j\rangle
= & \gamma_\mathrm{eff}
\left(-1+\xi_j\right)\langle a_j\rangle - \sqrt{\gamma} \langle a_{j,\mathrm{in}}\rangle \notag\\
& - \gamma_\mathrm{eff}
\left(\frac{\i\Lambda+\mathcal{C} e^{-\i\theta}}{2} \langle a_{j+1}\rangle + 
\frac{\i\Lambda+\mathcal{C} e^{\i\theta}}{2} \langle a_{j-1}\rangle \right) \notag \\
& \equiv \sum_\ell H_{j,\ell} \langle a_\ell\rangle - \sqrt{\gamma}\langle a_{j,\mathrm{in}}\rangle\,, \label{eq:eomDisorder}
\end{align}
with $\mathcal{C}\equiv2\Gamma/(\gamma+2\Gamma-\kappa)$, $\Lambda\equiv4J/(\gamma+2\Gamma-\kappa)$ and the effective local decay rate $\gamma_\mathrm{eff}\equiv(\gamma+2\Gamma-\kappa)/2$.
For future convenience, we also define the quantities $\mu_0\equiv-\gamma_\mathrm{eff}$, $\mu_\pm\equiv-\gamma_\mathrm{eff}(\i\Lambda+\mathcal{C}e^{\mp\i\theta})/2$.
In Eq.~\eqref{eq:eomDisorder} we introduced the central element of our analysis: we included disorder in terms of independent and identically distributed complex-valued random variables $\bm \xi=\mathrm{diag}(\xi_1,\dots,\xi_N)$.
The real part of this complex `potential' describes disorder in the on-site decay rates (rate disorder), which can either stem from the local loss or the incoherent pump rate, while the imaginary part accounts for disorder in the cavity frequencies (frequency disorder).

The system exhibits directional end-to-end amplification, under \emph{open boundary conditions} (OBC), for $\mathcal{C}^2\sin^2\theta>1$, which coincides with the regime of a non-trivial winding number of the dynamic matrix $\bm H$ under \emph{periodic boundary conditions} (PBC)~\cite{Wanjura2020}.
Its transport properties are encoded in the scattering matrix
\begin{align}
\bm S(\omega)
&
\equiv \mathbb{1} + \gamma (\i\omega\mathbb{1} + \bm H)^{-1}
\equiv \mathbb{1} + \gamma \bm M^{-1}(\omega)
\equiv \mathbb{1} + \gamma \bm \chi(\omega), \label{eq:SmatChimat}
\end{align}
where we defined the susceptibility matrix $\bm\chi\equiv \bm M^{-1}(\omega)$. The element $S_{j,\ell}$ relates a weak coherent input at cavity $\ell$ to the output at cavity $j$. Non-reciprocity occurs when $\vert\bm S\rvert$ is not symmetric, $\lvert S_{j,\ell}\rvert\neq \lvert S_{\ell,j}\rvert$, while amplification when ${\lvert S_{j,\ell}\rvert>1}$.
In non-trivial topological regimes, the scattering matrix features directional end-to-end forward gain $\mathcal{G}\gg 1$
\begin{align}
\mathcal{G} & \equiv
      \begin{cases}
         \lvert S_{1,N}\rvert^2 = \gamma^2\lvert\chi_{1,N}\rvert^2 & : \pi < \theta < 2\pi \ (\nu=+1) \\
         \lvert S_{N,1}\rvert^2 = \gamma^2\lvert\chi_{N,1}\rvert^2 & : 0 < \theta < \pi \ (\nu=-1)
      \end{cases} \label{eq:gain}
\end{align}
which grows \emph{exponentially} with system size, while the end-to-end reverse gain $\bar{\mathcal{G}}$, i.e.~the transmission in the reverse direction
\begin{align}
   \bar{\mathcal{G}} & \equiv
      \begin{cases}
         \lvert S_{N,1}\rvert^2 = \gamma^2\lvert\chi_{N,1}\rvert^2 & : \pi < \theta < 2\pi \ (\nu=+1) \\
         \lvert S_{1,N}\rvert^2 = \gamma^2\lvert\chi_{1,N}\rvert^2 & : 0 < \theta < \pi \ (\nu=-1),
      \end{cases} \label{eq:revGain}
\end{align}
is exponentially suppressed with system size. From the last equality in~\eqref{eq:SmatChimat} it is clear that the relevant dynamical features are encoded in the susceptibility matrix $\bm\chi$. In the following, we thus focus on $\bm\chi$ and restrict our attention to the resonant response of the system, i.e., we set $\omega=0$.

\begin{figure}[t]
\centering
\includegraphics[width=0.5\textwidth]{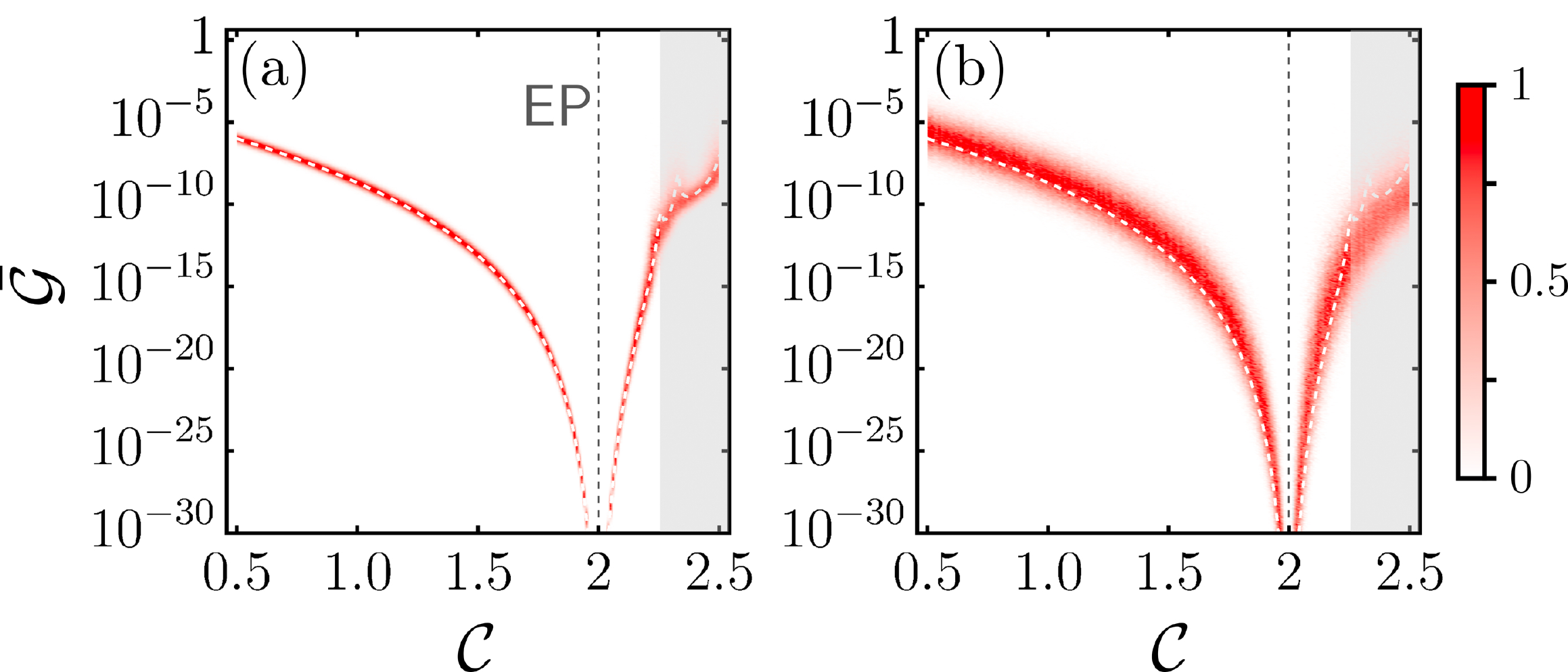}
\caption{\textbf{End-to-end reverse gain for uniform disorder.}
Even strong disorder has a very weak effect on the distribution of the reverse gain in the vicinity of the EP which is very narrow; at the EP the reverse gain is exactly zero.
Rate disorder with (a)~$w=0.25$ and (b)~$w=1.0$. Here, $N=10$, $\Lambda=2$, $\theta=\frac{\pi}{2}$. The gray area is the dynamically unstable regime. Dashed, white lines indicate the numerically calculated mean value which practically coinsides with the disorderless value.}
\label{fig:reverseGain}
\end{figure}

\textit{Perfect isolation in the presence of disorder.---}%
Without disorder, the choice $J=\Gamma/2$ (equivalently expressed as $\mathcal{C}=\Lambda$) and $\theta=\frac{\pi}{2},\frac{3\pi}{2}$ guarantees perfect one-way propagation, i.e., the reverse gain $\bar{\mathcal{G}}$ identically vanishes. This condition corresponds to an EP of the dynamic matrix~\cite{SM} and represents the optimal working point for directional amplifiers. For this case, we now show that $\bar{\mathcal{G}}$ remains \emph{exactly} zero under \emph{arbitrarily strong} disorder.
To see this analytically, we separate the diagonal contributions $\bm D=\mu_0\mathbb{1}+\bm\xi$ in the dynamic matrix $\bm H$ from the off-diagonals $\bm R=\bm H-\bm D$.
We use an expression derived from the Woodbury matrix identity~\cite{Press2007} to obtain $\bm\chi$ under OBC
\begin{align}
\bm\chi & = \sum_{n=0}^{N-1} (-1)^n (\bm R \bm D^{-1})^n \bm D^{-1}. \label{eq:directionalityExpansion}
\end{align}
We note that at the EP, $\bm R \bm D^{-1}$ is nilpotent, $(\bm R \bm D^{-1})^N=0$, so the sum only runs up to $N-1$.
At the EP, $\bm R$ is a strictly lower (upper) triangular matrix for $\theta=\frac{\pi}{2}$ ($\theta=\frac{3\pi}{2}$). Raising a strictly lower (upper) triangular matrix to the power $n$ yields again a lower (upper) triangular matrix, and each multiplication shifts the non-zero contributions further to the left (right) until $(\bm R \bm D^{-1})^N=0$.
Hence the other upper (lower) triangle remains \emph{exactly} zero independent of any on-site disorder.
In the neighborhood of the EP, $\bm R$ is not exactly lower (upper) triangular, but close to and the other triangle gets smaller and smaller with the order $n$ in the expansion~\eqref{eq:directionalityExpansion}, therefore still leading to an exponentially small reverse gain. We stress that the result makes no assumptions on the disorder distribution. 
In the relevant topologically non-trivial regime, isolation is always ideal (for $\Lambda=\mathcal{C}>1$) or exponentially close to (for $\Lambda\approx\mathcal{C}>1$), no matter how strong the disorder; the region of exponentially suppressed reverse gain gets larger with the number of cavities in the array $N$. These results are confirmed by inspecting Fig.~\ref{fig:reverseGain}, which display the distribution of $\bar{\mathcal{G}}$ at different disorder strengths $w$ (for uniform disorder), which is particularly narrow in the vicinity of the EP. It also provides a clear indication that, even further away from the EP, the non-reciprocal character of NH topological regimes is not disrupted by disorder ($\bar{\mathcal{G}}$ is highly suppressed and the distribution of sampled values remains narrow), as we will discuss in the following.

\begin{figure}[t]
\centering
\includegraphics[width=\linewidth]{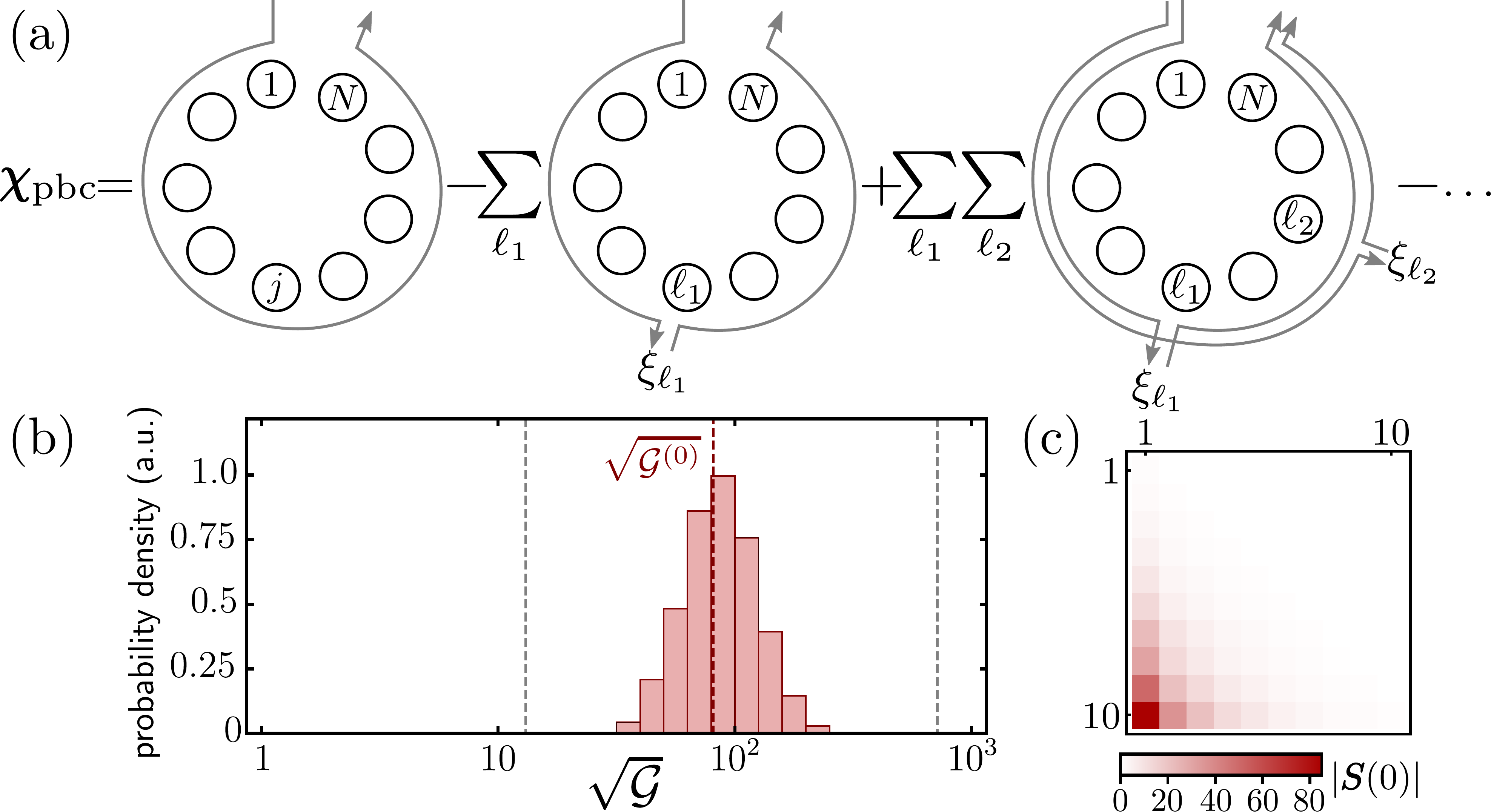}
\caption{\textbf{Expansion of the susceptibility matrix under PBC and distribution of the gain.}
(a)~In the presence of disorder, the susceptibility matrix $\bm\chi_\mathrm{pbc}$ may be written as an expansion summing over all possible paths of excitations scattering off sites $\{\ell_1,\dots,\ell_m\}$, see Eq.~\protect{\eqref{eq:expansionPBC}}. This expansion determines the susceptibility matrix under open boundary conditions, Eq.~\protect{\eqref{eq:expansionOBC0}}. 
(b)~The scattering matrix elements are randomly distributed in the presence of disorder. Here we show the distribution of the gain under uniform rate on-site disorder with $w=0.25$ which is centered around its disorderless value $\sqrt{\mathcal{G}^{(0)}}$, as well as (c) a representative example for the susceptibility matrix $\bm\chi_\mathrm{obc}$ related to the $S$-matrix according to Eq.~\protect{\eqref{eq:SmatChimat}}. The gray dashed lines indicate the bounds of the distribution~\protect{\eqref{eq:boundGain}}.
$\Lambda=2$, $\mathcal{C}=1.8$, $\theta=\frac{\pi}{2}$, $N=10$.}
\label{fig:PBCExpIntuitive}
\end{figure}

\textit{Topological correspondence in the presence of disorder.---}%
We now address the effect of disorder on \emph{all} $\bm\chi$ elements, and without restricting the analysis to a neighborhood of the EP.
We quantify the influence of disorder by expressing the matrix elements of $\bm\chi$ with disorder, in terms of those (known) without disorder. 
In order to do that, we assume that the probability distribution of $\xi_j$ has compact support, i.e. $\lvert\xi_j\rvert \leq w$. Directional amplification appears as we move from PBC to OBC, so we need to distinguish between $\bm\chi_\mathrm{pbc}$ and $\bm\chi_\mathrm{obc}$ (in the previous section $\bm\chi_\mathrm{obc}$ was simply denoted as $\bm\chi$). Our starting point is the susceptibility matrix $\bm\chi_\mathrm{pbc}$ under PBC in the presence of disorder.
According to Eq.~\eqref{eq:SmatChimat}, we have
$\bm\chi_\mathrm{pbc} = (\bm M^{(0)}_\mathrm{pbc} + \gamma_\mathrm{eff} \bm{\xi})^{-1}$,
where the index $(0)$ denotes quantities calculated without disorder, so that $\bm\chi^{(0)}_\mathrm{pbc}=[\bm M^{(0)}_\mathrm{pbc}]^{-1}$.
We write $\bm\chi_\mathrm{pbc}$ as a power series
again using the expansion derived from the Woodbury matrix identity
\begin{align}
\bm\chi_\mathrm{pbc} & = \sum_{n=0}^\infty (-1)^n \big(\bm\chi_\mathrm{pbc}^{(0)} \, \gamma_\mathrm{eff} \bm\xi\big)^n \bm\chi_\mathrm{pbc}^{(0)},
\end{align}
which can be written as
\begin{align}
\label{eq:expansionPBC}
(\bm \chi_\mathrm{pbc})_{j,\ell} = \chi^{(0)}_{j,\ell}
& - \sum_{\ell_1} \chi^{(0)}_{j,\ell_1} \gamma_\mathrm{eff}
\, \xi_{\ell_1} \chi^{(0)}_{\ell_1,\ell} \\
& + \sum_{\ell_1,\ell_2} \chi^{(0)}_{j,\ell_1} \gamma_\mathrm{eff} \, \xi_{\ell_1} \chi^{(0)}_{\ell_1,\ell_2} \gamma_\mathrm{eff} \, \xi_{\ell_2} \chi^{(0)}_{\ell_2,\ell}
+ \dots \notag
\end{align}
For brevity, we denote the matrix elements of $\bm\chi_\mathrm{pbc}^{(0)}$ by $\chi^{(0)}_{j,\ell}$.
This expansion may be interpreted as matrix Taylor expansion of $(\bm M^{(0)}_\mathrm{pbc} + \gamma_\mathrm{eff} \bm{\xi})^{-1}$ around $\bm\xi=0$,
which converges for
\begin{align}
   \max_j \, \lvert \xi_j\rvert
   & \leq
   \frac{\max_{j,\ell} \lvert (M^{(0)})_{j,\ell}\rvert}{\gamma_\mathrm{eff}}
   = \frac{\lvert \i\Lambda+\mathcal{C}e^{\mp\i\theta}\rvert}{2}. \label{eq:convergenceRadius}
\end{align}
The expansion (\ref{eq:expansionPBC}) offers the intuitive interpretation illustrated in Fig.~\ref{fig:PBCExpIntuitive}~(a).
$\bm\chi_\mathrm{pbc}$ is the sum over all possible paths of excitations through the system via sites $\ell_j$.
Each multiplication with $\chi_{\ell_{j-1},\ell_j}^{(0)}\xi_{\ell_j}$ indicates a path from site $\ell_{j-1}$ to $\ell_j$, with the transition induced by the disorder $\gamma_\mathrm{eff}\xi_{\ell_j}$.

Without disorder, a non-trivial topological regime implies non-reciprocal dynamics, i.e., an asymmetric $\bm\chi_\mathrm{pbc}^{(0)}$.  
Since each order in the sum~\eqref{eq:expansionPBC} involves a product of asymmetric matrices $\bm\chi_\mathrm{pbc}^{(0)} \bm\xi$, we conclude that non-reciprocity survives in the presence of disorder.
Moreover, a non-trivial winding number implies that some eigenvalues have positive real part, which indicates unstable dynamics~\cite{Wanjura2020}.
Since individual terms of Eq.~\eqref{eq:expansionPBC} contain $\bm\chi_\mathrm{pbc}^{(0)}$, this still occurs in the presence of disorder.
However, Eq.~\eqref{eq:expansionPBC} is a coherent superposition of these scattering processes via the disordered sites $j$, so depending on the phase of $\xi_j$, individual path segments may interfere destructively. Indeed, this leads to the main difference between rate and frequency disorder which we elaborate on below.
\begin{figure}[t]
\centering
\includegraphics[width=\columnwidth]{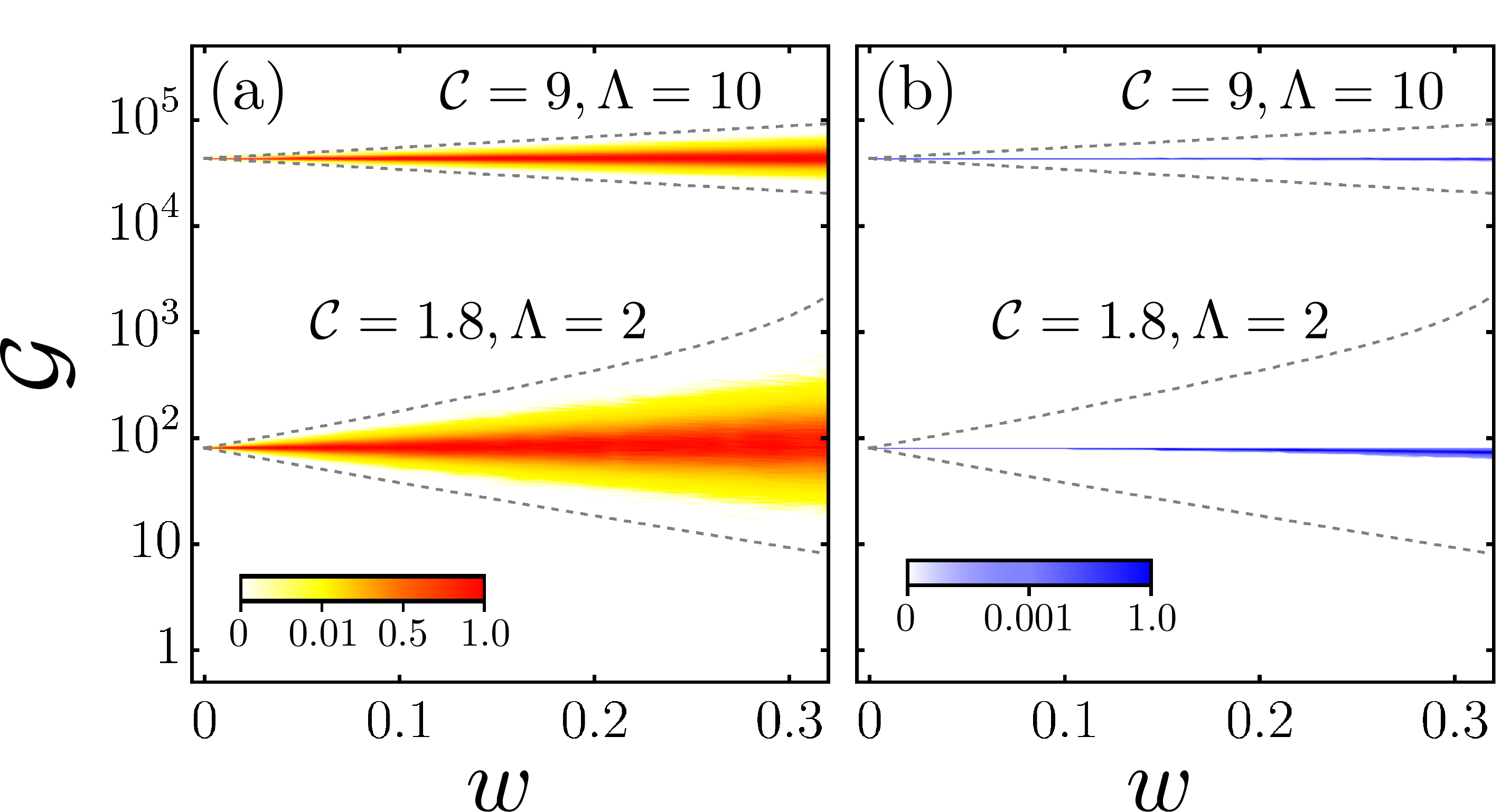}
\caption{\textbf{End-to-end gain for uniform disorder.}
Inequality~\protect{\eqref{eq:boundGain}} bounds the distribution of the gain (dashed lines). All sampled values (distribution indicated by color gradients) lie within this area.
Rate (a) and frequency disorder (b) influence the gain differently:
rate disorder leads to a broader distribution which may even enhance the gain, while frequency disorder leads to a narrow distribution with a slight average decrease.
Larger couplings $\mathcal{C}$, $\Lambda$
lead to tighter bounds than smaller couplings.
The bound shown in (b) is the same as in (a).
The peak of the distribution at each $w$ is normalized to $1$.
$\theta=\frac{\pi}{2}$ and $N=10$.
}
\label{fig:gainBounds}
\end{figure}

We now consider the system under OBC.
We recall that, in the disorderless case, moving to OBC leads to stable dynamics and exponential directional end-to-end gain (for $\nu=\pm1$), i.e., the matrix $\bm \chi_\mathrm{obc}^{(0)}$ is dominated by a single off-diagonal corner element.
We perform the change in boundary conditions by removing the matrix corners $\bm M_\mathrm{obc}=\bm M_\mathrm{pbc} - (\mu_+ \ketbra{N}{1} + \mu_- \ketbra{1}{N})$. Using the inversion formula from Ref.~\cite{Miller1981} we obtain
\begin{align}
   \bm\chi_\mathrm{obc}
      = &
      \frac{\mu_{-}}{1+g_-} \bm\chi_\mathrm{pbc} \ketbra{1}{N} \bm\chi_\mathrm{pbc}
      +
      \frac{\mu_{+}}{1+g_+} \bm\chi_\mathrm{pbc} \ketbra{N}{1} \bm\chi_\mathrm{pbc} \notag \\
      & + \bm\chi_\mathrm{pbc}
      + \mathcal{O}(c^N)\,, \label{eq:expansionOBC0}
\end{align}
with $g_+=\mu_+(\chi_\mathrm{pbc})_{1,N}$, $g_-=\mu_-(\chi_\mathrm{pbc})_{N,1}$, and $\lvert c\rvert<1$ indicating an exponentially small correction. We use Dirac notation to denote matrix elements in the site basis $\{\ket{j}\}$.

Eq.~\eqref{eq:expansionOBC0} relates the susceptibility matrix of the disordered open chain to $\bm\chi_\mathrm{pbc}$, which is in turn connected to the disorderless case via expansion~\eqref{eq:expansionPBC}. This result allows us to evaluate the effects of disorder in NH topological regimes. While the physical consequences of this expression will become clear in the next section, we briefly comment on the structure of Eq.~\eqref{eq:expansionOBC0}.
The first (second) term is the dominant term for $\nu=-1$ ($\nu=+1$).
The asymmetry (non-reciprocity) is due to both asymmetric $\bm\chi_\mathrm{pbc}$ and corner elements $\ketbra{1(N)}{N(1)}$, while
the exponentially large gain factor is determined by the pre-factor $\mu_\pm/(1+g_\pm)$, as we will see next. 

\textit{Placing bounds on the distribution of the gain.---}%
Once the probability distribution of the disorder is known, Eq.~\eqref{eq:expansionOBC0} in principle allows to compute the distribution of $\bm\chi_\mathrm{obc}$ elements.
However, obtaining the exact distribution or the exact expression of the moments becomes an intractable problem even for simple distributions of the disorder, e.g. uniform disorder. Instead, it is possible to derive bounds for the probability distribution of $\bm\chi_\mathrm{obc}$, using our \emph{only} assumption $\lvert \xi_j\rvert<w$.
We first bound the pre-factor of Eq.~\eqref{eq:expansionOBC0}, which determines the gain, and obtain
\begin{align}
   \left\lvert\lvert\varepsilon_{\nu(1-N)}\rvert - \rho^\mathrm{min}_{j,\ell}\right\rvert
   \leq
   \left\lvert \frac{1 + g_\pm}{\mu_\pm}\right\rvert
   \leq
   \lvert\varepsilon_{\nu(1-N)}\rvert + \rho^\mathrm{max}_{j,\ell}\,,  \label{eq:boundsDenominator}
\end{align}
in which $j=1,\ell=N$ for $\nu=+1$ and vice versa for $\nu=-1$. (see SM for the derivation~\cite{SM}). We introduced the quantities 
\begin{align}
   \rho^\mathrm{max}_{j,\ell}
      & = \sum_{n=1}^\infty \big(\big\lvert\chi^{(0)}\big\rvert^{n+1}\big)_{j,\ell} \gamma_\mathrm{eff}^n w^n, \\
   \rho^\mathrm{min}_{j,\ell}
      & = \left\lvert\sum_{n=1}^\infty (-1)^n \big(\big[\chi^{(0)}\big]^{n+1}\big)_{j,\ell} \gamma_\mathrm{eff}^n w^n \right\rvert,
\end{align}
which may be calculated numerically or estimated analytically, see SM~\cite{SM}. Notice that, for $w\rightarrow0$, both  $\rho^\mathrm{min/max}_{j,\ell}$ vanish and Eq.~\eqref{eq:boundsDenominator} reduces to $\varepsilon_{\nu(1-N)}=(1+g_\pm^{(0)})/\mu_\pm$. In the SM we also provide a geometrical interpretation of these bounds and discuss the different effects of rate and frequency disorder.

In a second step, using a similar approach, we derive bounds for \emph{all the susceptibility matrix elements}  (see SM~\cite{SM}). Here, we focus on the end-to-end forward gain
\begin{align}
   \frac{\left\lvert\lvert \chi^{(0)}_{j,\ell}\rvert - \rho^\mathrm{min}_{j,\ell}\right\rvert^2}{\lvert \varepsilon_{\nu(1-N)}\rvert + \rho^\mathrm{max}_{j,\ell}}
   \leq
   \sqrt{\mathcal{G}}
   \leq
   \frac{\left(\lvert \chi^{(0)}_{j,\ell}\rvert + \rho^\mathrm{max}_{j,\ell}\right)^2}{\left\lvert\lvert \varepsilon_{\nu(1-N)}\rvert - \rho^\mathrm{min}_{j,\ell}\right\rvert}, \label{eq:boundGain}
\end{align}
with $j=1,\ell=N$ for $\nu=+1$ and vice versa for $\nu=-1$.
Inequality~\eqref{eq:boundGain} shows us that the large gain in non-trivial topological regimes is at most modified by $\rho^\mathrm{min/max}_{j,\ell}$ by the disorder. In Fig.~\ref{fig:PBCExpIntuitive}~(b) we show the bounds~\eqref{eq:boundGain} for a sampled distribution of the gain. The distribution peaks close to its original value without disorder $\sqrt{\mathcal{G}^{(0)}}=\lvert\chi^{(0)}_{j,\ell}/\varepsilon_{\nu(1-N)}\rvert$, but is broadened by the (rate) disorder. In particular, the distribution has a long tail, yet it has compact support within the bounds~\eqref{eq:boundGain}.
We assumed uniform on-site disorder in Fig.~\ref{fig:PBCExpIntuitive}~(b), but any on-site disorder with compact support would be bounded in the same way.
We also show an instance of $\bm\chi_\mathrm{obc}$ in Fig.~\ref{fig:PBCExpIntuitive}~(c) for a single representative realization of rate disorder, from which directional amplification is apparent.

An analytic estimate of $\rho^\mathrm{min/max}_{j,\ell}$ shows that the bounds do \emph{not} change the exponential scaling of the gain with $N$~\cite{SM}.
This conclusion holds even in the thermodynamic limit and proves that the defining feature of NH topological amplification, i.e. its exponential scaling, is robust against disorder.
We stress that the analytic bounds we have derived are essential to draw this conclusion, since extracting the scaling by sampling numerically from Eq.~\eqref{eq:expansionOBC0} would have been unfeasible ($\bm\chi_\mathrm{obc}$ becomes ill-conditioned already for $N\approx20$). 

We illustrate the bounds~\eqref{eq:boundGain} in Fig.~\ref{fig:gainBounds}, as a function of the disorder strength $w$, together with the sampled distribution of the gain for uniform rate disorder $\xi_j\in[-w,w]$ or frequency disorder $\i \xi_j\in[-w,w]$ and for different coupling strengths between neighboring sites; the distribution in Fig.~\ref{fig:PBCExpIntuitive}~(b) is one slice of this plot taken at $w=0.25$.
Fig.~\ref{fig:gainBounds} clearly shows the difference between rate and frequency disorder. The former leads to a broader distribution of the gain and can even lead to its increase. For sufficiently large $w$, it can induce a dynamic instability when the disorder in the pump rates overcomes the local damping.
While we have adjusted the color scheme in Fig.~\ref{fig:gainBounds} to reveal the full extent of the distribution, we note that it is actually quite narrow---even more so for frequency disorder---and peaks close to the original value of the gain even for large $w$.
Rate disorder preserves the phase in each summand of the coherent expansion Eq.~\eqref{eq:expansionPBC}, and therefore the disorder directly affects the absolute value of $\bm\chi$ making the distribution of the gain broader. In contrast, frequency disorder changes the phase of each summand in Eq.~\eqref{eq:expansionPBC} leaving the absolute value of $\bm\chi$ almost unchanged which leads to a narrow distribution of the gain.
Furthermore, Fig.~\ref{fig:gainBounds} illustrates that a larger cooperativity $\mathcal{C}$ or hopping constant $\Lambda$ leads to narrower bounds, which can be exploited in practical applications.

Finally, we stress two more merits of our approach: (i) The fact that we bound the full distribution of the scattering matrix entails that our treatment is valid for \emph{any realization} of the disorder, even single instances. Our conclusions do not rely on computing statistical moments and automatically account for rare events. For practical purposes, this feature seems especially appealing. (ii) Our results hold true for \emph{any} probability distribution with compact support. The latter assumption \emph{per se} does not guarantee the existence of the bounds, since even distributions with compact support may lead---via the matrix inverse in~\eqref{eq:SmatChimat}---to distributions whose domain is not bounded, e.g.~in the dynamically unstable regime.

\textit{Conclusions.---}%
We showed that non-Hermitian topological amplification in coupled cavity arrays is robust against complex disorder, which opens the doors to robust NH topological amplifiers and sensors.
From the practical point of view, it shows the feasibility of directional amplifiers, e.g.~to read out fragile quantum signals in superconducting quantum devices or for applications in quantum metrology. From the theory point of view, our original approach---based on the scattering matrix and placing bounds on its distribution---can find applications beyond topological amplification, e.g.~the study of disorder-induced localization in NH coupled-cavity arrays.

\begin{acknowledgements}
\textit{Acknowledgements.---}
C.C.W. acknowledges the funding received from the Winton Programme for the Physics of Sustainability and EPSRC (Project Reference EP/R513180/1).
A.N. holds a University Research Fellowship from the Royal Society and acknowledges additional support from the Winton Programme for the Physics of Sustainability.
We acknowledge the funding received from the European Union's Horizon 2020 research and innovation programme under Grant No.~732894 (FET Proactive HOT).
\end{acknowledgements}


%


\clearpage
\appendix
\twocolumngrid
\begin{center}
\widetext
{\large\bf Correspondence between non-Hermitian topology and directional amplification in the presence of disorder\\---Supplementary Material---} \\[\baselineskip]
{\normalsize
Clara C. Wanjura, Matteo Brunelli, and Andreas Nunnenkamp}
\end{center}

\setcounter{equation}{0}
\setcounter{figure}{0}
\setcounter{table}{0}
\setcounter{page}{1}

\renewcommand{\theequation}{S\arabic{equation}} 
\renewcommand{\thefigure}{S\arabic{figure}}

\begin{center}
\textbf{Determining the exceptional point (EP)}
\end{center}
The value of the EP without disorder can be extracted analytically for all $N$. At the EP, eigenvalues and eigenvectors coalesce. The dynamic matrix of the disorderless version of the system~\eqref{eq:eomDisorder} is a Toeplitz matrix, for which there exists an analytic expression for both eigenvalues and eigenvectors~\cite{Willms2008}
\begin{align}
   \lambda_m
      =
          & \i\omega +
          \frac{\gamma_\mathrm{eff}}{2}
          \left[\vphantom{\left[\frac{m\pi}{N+1}\right]}
             -1 \right.
             + \left.
             \sqrt{(\mathcal{C} e^{\i\theta} + \i\Lambda)
                   (\mathcal{C} e^{-\i\theta} + \i\Lambda)
             }
             \cos \left(\frac{m\pi}{N+1}\right)
          \right]. \label{eq:eigenvals}
\end{align}
From this expression it is clear, that the eigenvalues can only coalesce when either
$\mathrm{i} \Lambda=-e^{\mathrm{i}\theta}\mathcal{C}$
or
$\mathrm{i} \Lambda=-e^{-\mathrm{i}\theta}\mathcal{C}$,
in which case the dynamic matrix becomes an upper (lower) triangular matrix with only the diagonal and super-(sub-)diagonal non-zero. Since all the entries on the respective diagonal and super-(sub-)diagonal are the same, the matrix has rank $1$ and these are indeed exceptional points. We obtain the $N$-fold degenerate right eigenvectors from Gaussian elimination to be either $(1,0,\dots,0,0)^\mathrm{T}$ in the former case or $(0,0,\dots,0,1)^\mathrm{T}$ in the latter case.

Whenever we refer to the EP, we mean the position of the EP without disorder.
The disorder may remove the exceptional point, but this is of no consequence for our results.

\begin{center}
\textbf{Bounds for the susceptibility matrix elements}
\end{center}
\noindent\textbf{Derivation of the bounds for the susceptibility matrix elements:}
Here, we derive the bounds~\eqref{eq:boundsDenominator} and \eqref{eq:boundGain} as well as bounds for the other susceptibility matrix elements.
We obtain the bounds from the expansion of Eq.~\eqref{eq:expansionOBC0}, which, for convenience, is repeated below
\begin{align}
   \bm\chi_\mathrm{obc}
      = &
      \underbrace{
      \frac{\mu_{-}}{1+g_-} \bm\chi_\mathrm{pbc} \ketbra{1}{N} \bm\chi_\mathrm{pbc}
      }_{\equiv \Sigma_-}
      +
      \underbrace{
      \frac{\mu_{+}}{1+g_+} \bm\chi_\mathrm{pbc} \ketbra{N}{1} \bm\chi_\mathrm{pbc}
      }_{\equiv \Sigma_+}
      + \bm\chi_\mathrm{pbc}
      + \mathcal{O}(c^N) \tag{\ref{eq:expansionOBC0}}
\end{align}
with $g_+=\mu_+(\chi_\mathrm{pbc})_{1,N}$, $g_-=\mu_-(\chi_\mathrm{pbc})_{N,1}$, and an exponentially small correction $\mathcal{O}(c^N)$ with $\lvert c\rvert<1$ for $\nu\neq0$. Without disorder, we saw that depending on $\nu$ either the second or the third summand dominates and leads to directional end-to-end gain.

We focus on the case $\nu=-1$ ($\nu=+1$ follows analogously exchanging the indices $N$ and $1$ in the expressions below), for which $\Sigma_-$ yields the dominant contribution to the gain, and obtain the upper and lower bound on the absolute value of its elements.
We explicitly write down the expansion for $\Sigma_-$ using Eq.~\eqref{eq:expansionPBC} in which we have set $\gamma_\mathrm{eff}=1$ in this section
\begin{align}
   \Sigma_-
   =
   \frac{\mu_-}{1+g_-}
   \sum_{j,\ell}\sum_{m,n=0}^\infty \sum_{\substack{ r_1,\dots,r_n \\ s_1,\dots,s_m}} (-1)^{n+m}
   &
   \chi^{(0)}_{j,r_1} \chi^{(0)}_{r_1,r_2} \dots \chi^{(0)}_{r_{n-1},r_n} \chi^{(0)}_{r_n,1}
   \chi^{(0)}_{N,s_1} \chi^{(0)}_{s_1,s_2} \dots \chi^{(0)}_{s_{n-1},s_n} \chi^{(0)}_{s_m,\ell} \notag \\
   & \xi_{r_1} \xi_{r_2} \dots \xi_{r_n}
   \xi_{s_1} \xi_{s_2} \dots \xi_{s_m} \ketbra{j}{\ell} \label{eq:expansionOBC}.
\end{align}
We see that the product inside the sum is the product of two copies of the expressions~\eqref{eq:expansionPBC}.
However, the change of boundary conditions has added scattering processes to the first site and scattering processes originating from the $N$th site compared to expansion~\eqref{eq:expansionPBC}.

To obtain the lower and upper bound, we consider the pre-factor $\mu_-/(1+g_-)$ and the sum separately.
First, we have a closer look at the pre-factor which sets the magnitude of the gain. The pre-factor also limits the validity range of the expansion since sufficiently strong rate (real) disorder may induce a dynamic instability which coincides with the convergence radius~\eqref{eq:convergenceRadius}.
We expand
\begin{align*}
   1+g_- & = 1 - \mu_- (\chi_\mathrm{pbc})_{N,1} = 1 - \mu_- \sum_{n=0}^\infty \sum_{\ell_1,\dots,\ell_n} (-1)^n  \chi^{(0)}_{N,\ell_1} \chi^{(0)}_{\ell_1,\ell_2} \dots \chi^{(0)}_{\ell_{n-1},\ell_n} \chi^{(0)}_{\ell_n,1} \xi_{\ell_1} \xi_{\ell_2} \dots \xi_{\ell_n}.
\end{align*}
We rewrite this expression recalling that $1 - \mu_- \chi_{N,1}^{(0)} = \mu_- \varepsilon_{\nu(1-N)}$ is exponentially small (see main text and~\cite{Wanjura2020})
\begin{align*}
   1 + g_- & = 1 - \mu_- \chi_{N,1}^{(0)} - \mu_- \sum_{n=1}^\infty \sum_{\ell_1,\dots,\ell_n} (-1)^n \chi^{(0)}_{N,\ell_1} \chi^{(0)}_{\ell_1,\ell_2} \dots \chi^{(0)}_{\ell_{n-1},\ell_n} \chi^{(0)}_{\ell_n,1} \xi_{\ell_1} \xi_{\ell_2} \dots \xi_{\ell_n} \\
   & = \mu_- \varepsilon_{\nu(1-N)} - \mu_- \sum_{n=1}^\infty \sum_{\ell_1,\dots,\ell_n} (-1)^n \chi^{(0)}_{N,\ell_1} \chi^{(0)}_{\ell_1,\ell_2} \dots \chi^{(0)}_{\ell_{n-1},\ell_n} \chi^{(0)}_{\ell_n,1} \xi_{\ell_1} \xi_{\ell_2} \dots \xi_{\ell_n}.
\end{align*}
Using the triangle inequality we obtain the estimate
\begin{align}
   \left\lvert \frac{1 + g_-}{\mu_-}\right\rvert
   & \leq \lvert \varepsilon_{\nu(1-N)}\rvert
   + \sum_{n=1}^\infty \sum_{\ell_1,\dots,\ell_n} \left\lvert\chi^{(0)}_{N,\ell_1} \chi^{(0)}_{\ell_1,\ell_2} \dots \chi^{(0)}_{\ell_{n-1},\ell_n} \chi^{(0)}_{\ell_n,1} \xi_{\ell_1} \xi_{\ell_2} \dots \xi_{\ell_n}\right\rvert \notag \\
   & \leq \lvert \varepsilon_{\nu(1-N)}\rvert
   + \sum_{n=1}^\infty \sum_{\ell_1,\dots,\ell_n} \left\lvert\chi^{(0)}_{N,\ell_1} \chi^{(0)}_{\ell_1,\ell_2} \dots \chi^{(0)}_{\ell_{n-1},\ell_n} \chi^{(0)}_{\ell_n,1}\right\rvert
   w^n \notag \\
   & = \lvert \varepsilon_{\nu(1-N)}\rvert
   + \sum_{n=1}^\infty \left\lvert\big(\big[\chi^{(0)}\big]^{n+1}\big)_{N,1}\right\rvert
   w^n \label{eq:upperBoundCalc}
\end{align}
noting that $\left\lvert\big(\big[\chi^{(0)}\big]^n\big)_{N,1}\right\rvert\ll1$ for $n\ll N$.
In a similar way, we obtain an upper bound from the triangle inequality
\begin{align}
   \left\lvert \frac{1 + g_-}{\mu_-}\right\rvert
   & \geq \left\lvert \lvert \varepsilon_{\nu(1-N)}\rvert
   - \left\lvert\sum_{n=1}^\infty \sum_{\ell_1,\dots,\ell_n} (-1)^n \chi^{(0)}_{N,\ell_1} \chi^{(0)}_{\ell_1,\ell_2} \dots \chi^{(0)}_{\ell_{n-1},\ell_n} \chi^{(0)}_{\ell_n,1} \xi_{\ell_1} \xi_{\ell_2} \dots \xi_{\ell_n}\right\rvert \right\rvert \notag \\
   & \geq
   \left\lvert \lvert \varepsilon_{\nu(1-N)}\rvert
   - \left\lvert\sum_{n=1}^\infty (-1)^n \big(\big[\chi^{(0)}\big]^{n+1}\big)_{N,1}
   w^n \right\rvert \right\rvert.
\end{align}
In the last step we have chosen $w$ small enough such that the second sum is smaller than $\lvert\varepsilon_{\nu(1-N)}\rvert$. This means that the estimate of that bound has a validity range which may be smaller than suggested by the convergence radius~\eqref{eq:convergenceRadius}.

\begin{figure}[htbp]
   \centering
   \includegraphics[width=.4\textwidth]{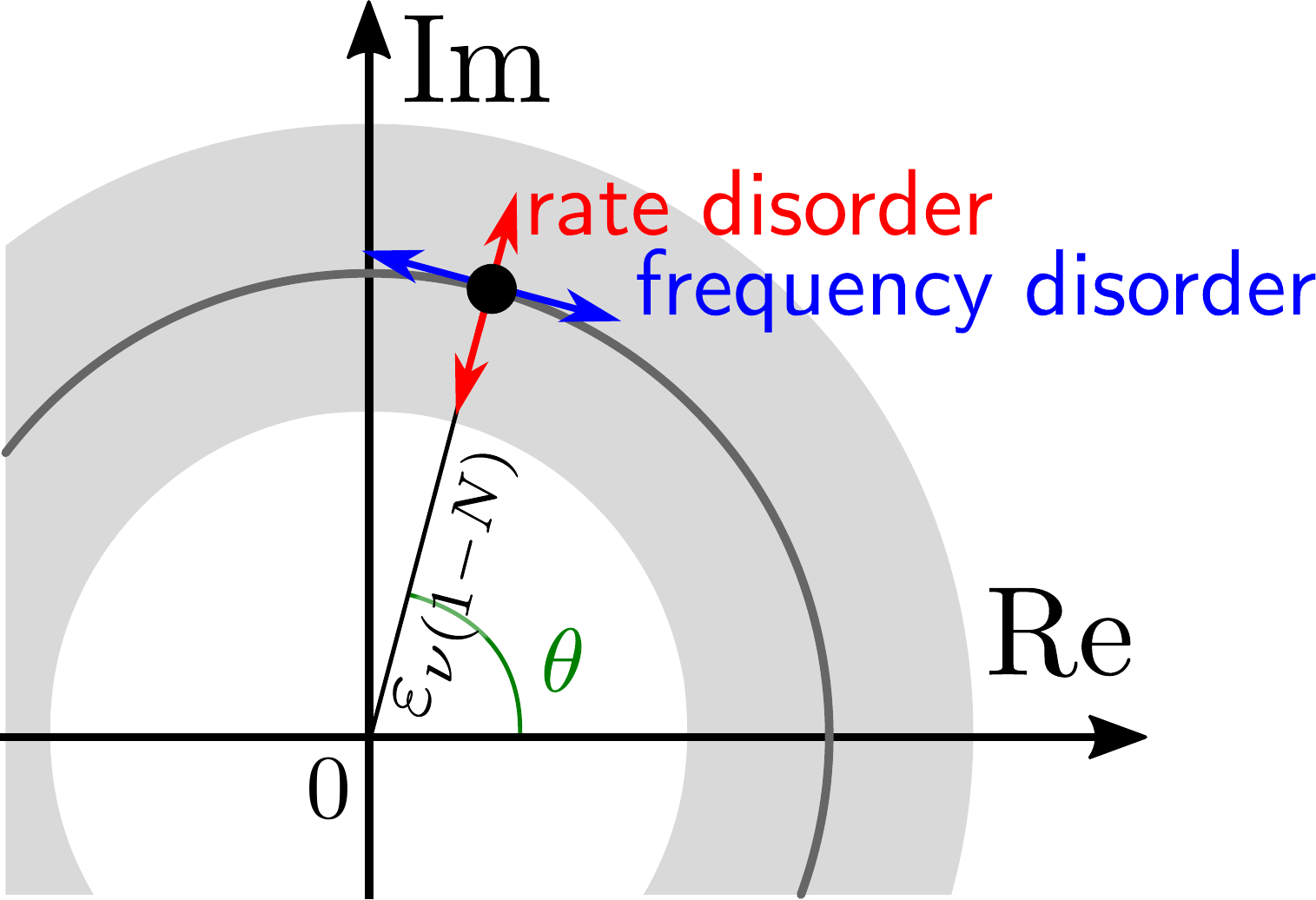}
   \caption{\textbf{Geometric interpretation of the bounds for the pre-factor determining the gain, Eq.~\protect{\eqref{eq:boundsDenominator}}.}
   The gain under OBC is crucially determined by the pre-factors $\mu_\pm/(1+g_\pm)$ in Eq.~\protect{\eqref{eq:expansionOBC0}} for which we show the bound (gray area) here. Different types of disorder (rate or frequency disorder) have different effects on this pre-factor. In particular, rate disorder can lead to gain enhancement as it can reduce the modulus of the pre-factor.
   }
   \label{fig:amplMechanism}
\end{figure}

Combining the upper and lower bound for the pre-factor, we find that the denominator can at most be modified by the terms
\begin{align}
   \rho^\mathrm{min}_{N,1} \equiv \left\lvert\sum_{n=1}^\infty (-1)^n \big(\big[\chi^{(0)}\big]^{n+1}\big)_{N,1}
      w^n \right\rvert,
   \quad\quad
   \rho^\mathrm{max}_{N,1} \equiv \sum_{n=1}^\infty \big(\big\lvert\chi^{(0)}\big\rvert^{n+1}\big)_{N,1}
      w^n \label{eq:rhoBounds}
\end{align}
such that we recover expression~\eqref{eq:boundsDenominator} of the main text
\begin{align}
   \left\lvert\lvert\varepsilon_{\nu(1-N)}\rvert - \rho^\mathrm{min}_{N,1}\right\rvert \leq
   \left\lvert \frac{1 + g_-}{\mu_-}\right\rvert \leq \lvert\varepsilon_{\nu(1-N)}\rvert + \rho^\mathrm{max}_{N,1}.  \tag{\ref{eq:boundsDenominator}}
\end{align}
This pre-factor $\lvert (1 + g_-)/\mu_-\rvert$ has a simple geometric interpretation as we argue below and in Fig.~\ref{fig:amplMechanism}.

Secondly, we analogously obtain the bounds for the sums in $\Sigma_-$ of Eq.~\eqref{eq:expansionOBC} and combine both expressions to obtain the bounds for all elements of $\Sigma_-$, which at the same time is an excellent approximation for the bounds of the elements of $\chi_\mathrm{obc}$.
We find for the dominant matrix corner
\begin{align}
   \frac{\left\lvert\lvert \chi^{(0)}_{N,1}\rvert - \rho^\mathrm{min}_{N,1}\right\rvert^2}{\lvert \varepsilon_{\nu(1-N)}\rvert + \rho^\mathrm{max}_{N,1}}
   \leq
   \lvert(\Sigma_-)_{N,1}\rvert
   \leq
   \frac{\left(\lvert \chi^{(0)}_{N,1}\rvert + \rho^\mathrm{max}_{N,1}\right)^2}{\left\lvert\lvert \varepsilon_{\nu(1-N)}\rvert - \rho^\mathrm{min}_{N,1}\right\rvert},
\end{align}
and in general for all elements $(\Sigma_+)_{j,\ell}$
\begin{align}
   \frac{\left\lvert\lvert \chi^{(0)}_{j,1}\rvert - \rho^\mathrm{min}_{j,1}\right\rvert \left\lvert\lvert \chi^{(0)}_{N,\ell}\rvert - \rho^\mathrm{min}_{N,\ell}\right\rvert}{\lvert \varepsilon_{\nu(1-N)}\rvert + \rho^\mathrm{max}_{1,N}}
   \leq
   \lvert(\Sigma_-)_{j,\ell}\rvert
   \leq
   \frac{\left(\lvert \chi^{(0)}_{j,1}\rvert + \rho^\mathrm{max}_{j,1}\right)\left(\lvert \chi^{(0)}_{N,\ell}\rvert + \rho^\mathrm{max}_{N,1}\right)}{\left\lvert\lvert \varepsilon_{\nu(1-N)}\rvert - \rho^\mathrm{min}_{1,N}\right\rvert}. \label{eq:boundsAllElements}
\end{align}
Using similar techniques as above, one can show that the contribution of $\Sigma_+$ is negligible in the topological non-trivial regime also in the presence of disorder. This implies that the bounds Eqs.~\eqref{eq:boundsAllElements} are in fact excellent estimates of the bounds for the susceptibility matrix elements $\lvert\chi_{j,\ell}\rvert$ and in particular of the gain, see expression~\eqref{eq:boundGain}.

We can either determine $\rho_{N,1}^\mathrm{min,max}$ numerically or approximate it analytically.
We provide the analytic estimate in the last section. 
Numerical evaluations of the bounds are computationally feasible for even very large system sizes, since the bounds are expressed in terms of the disorderless susceptibility matrix for which in principle there is even an analytic expression available~\cite{daFonseca2001}.
This is a great advantage of our analytical approach over purely numerical studies since large system sizes can be studied, whereas this would be numerically intractable as the dynamic matrix already becomes ill-conditioned for $N\approx20$. \\

\noindent\textbf{Geometric interpretation of the bounds:}
\noindent The bounds of the dominant pre-factor $(1+g_\pm)/\mu_\pm$, condition~\eqref{eq:boundsDenominator}, which determines the gain under OBC in Eq.~\eqref{eq:expansionOBC0}, are vital to derive the bounds for the scattering matrix elements. We repeat condition~\eqref{eq:boundsDenominator} here for completeness
\begin{align}
   \left\lvert\lvert\varepsilon_{\nu(1-N)}\rvert - \rho^\mathrm{min}_{j,\ell}\right\rvert
   \leq
   \left\lvert \frac{1 + g_\pm}{\mu_\pm}\right\rvert
   \leq
   \lvert\varepsilon_{\nu(1-N)}\rvert + \rho^\mathrm{max}_{j,\ell}  \tag{\ref{eq:boundsDenominator}}.
\end{align}
Examining condition~\eqref{eq:boundsDenominator} more closely, we uncover a geometric interpretation of this inequality which also highlights the crucial differences between rate and frequency disorder.
We illustrate condition~\eqref{eq:boundsDenominator} in Fig.~\ref{fig:amplMechanism}. The factor $(1+g_\pm)/\mu_\pm$ is bounded by the ring (gray area). The value of $(1+g_\pm)/\mu_\pm$ without disorder, $\varepsilon_{\nu(1-N)}$, lies between the two bounds set by the inner radius $\rho^\mathrm{min}_{j,\ell}$ and the outer radius $\rho^\mathrm{max}_{j,\ell}$ which both depend on the disorder strength $w$.
Depending on the type of the disorder (disorder in the rates or in the cavity frequencies), the data obtained from individual realizations distribute differently in the ring. This is indicated by the red and blue arrows in Fig.~\ref{fig:PBCExpIntuitive}~(b).

Rate disorder produces values of $(1+g_\pm)/\mu_+$ that lie on a normal to the circle denoting $\lvert\varepsilon_{\nu(1-N)}\rvert$. It has the angle $\theta$ with the real axis in Fig.~\ref{fig:PBCExpIntuitive}~(b). Therefore, the gain can decrease, or even increase, in individual realizations leading to a broad distribution, see for instance Fig.~\ref{fig:PBCExpIntuitive}~(c), while the phase of the susceptibility matrix element is (almost) unchanged.
Conversely, frequency disorder leads to values distributed tangentially to the circle of $\lvert\varepsilon_{\nu(1-N)}\rvert$, which leaves the absolute value of the gain (almost) unchanged while the phase of the susceptibility matrix element is subjected to stronger deviations.
The difference between rate and frequency disorder is also an expression of the coherent superposition of scattering events in expansion~\eqref{eq:expansionPBC}. Since rate disorder is real, it does not change the phase of any of the summands in Eq.~\eqref{eq:expansionPBC}. However, the disorder may strongly affect the magnitude of $(\chi_\mathrm{pbc})_{j,\ell}$, i.e. the gain. For frequency disorder, each summand in the sum~\eqref{eq:expansionPBC} comes with a random phase while the absolute value is less affected. This leads to a narrower distribution of the gain, but a slight decrease of the expected gain for stronger disorder through destructive interference. \\

\noindent\textbf{The bounds preserve the exponential scaling with system size:}
\begin{figure}[htbp]
   \centering
   \includegraphics[width=.8\textwidth]{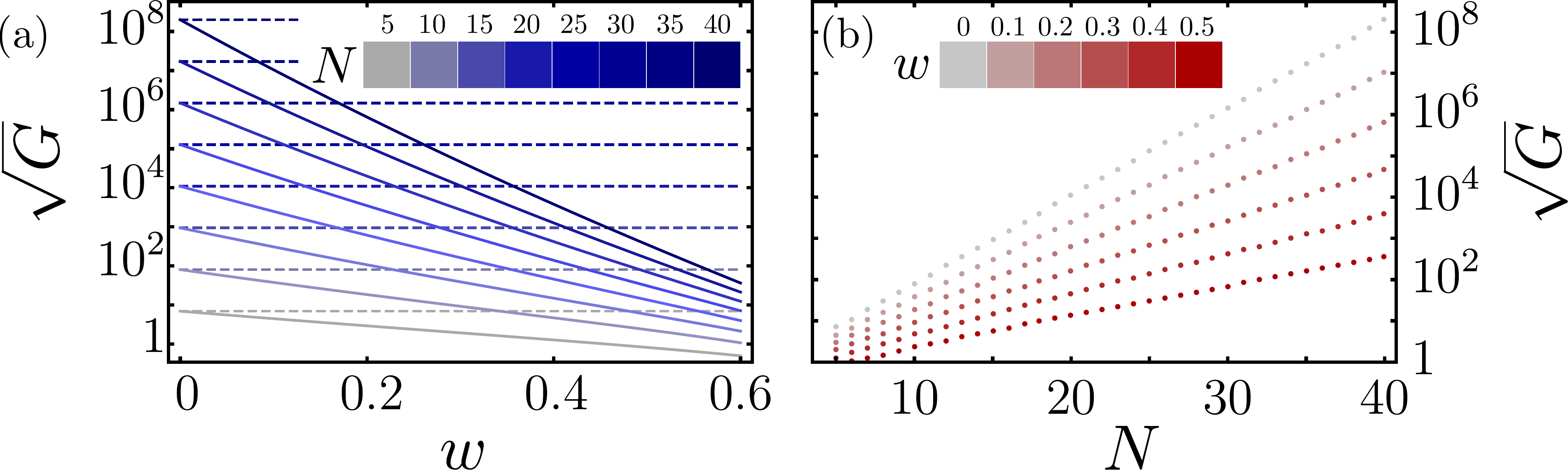}
   \caption{\textbf{Dependence of the bounds on system size and disorder strength.}
   (a)~The analytical lower bound, according to \protect{\eqref{eq:boundGain}}, for the gain (solid lines) as a function of the disorder strength $w$ at different chain lengths compared to the value without disorder (dashed lines).
   (b)~The lower bound for the gain as a function of system size for different $w$.
   The lower bound grows exponentially with system size although the slope in a logarithmic plot depends on $w$. This result extends to the thermodynamic limit and shows that even for $N\gg1$ the directional gain survives in the presence of sufficiently small disorder.
   $\mathcal{C}=1.8$, $\Lambda=2$, and $\theta=\frac{\pi}{2}$.
   }
   \label{fig:boudnsNw}
\end{figure}%
We show that the bounds on the gain, \eqref{eq:boundGain}, preserve the exponential scaling of the gain with system size. This holds true also in the thermodynamic limit $N\to\infty$, which sets these systems apart from Hermitian systems that are prone to Anderson localization.

We again focus on the case $\nu=-1$. The case $\nu=+1$ follows analogously exchanging the indices $N$ and $1$ below.
To preserve the exponential scaling, we need to ensure that the lower bound of the gain grows exponentially with system size.
We show this by proving that $\rho_{j,\ell}^\mathrm{max/min}$ decay exponentially with system size. Exponentially small $\rho^\mathrm{max}_{N,1}$ implies that the denominator $\varepsilon_{\nu(1-N)}+\rho^\mathrm{max}_{N,1}$ remains exponentially small (leading to an exponential growth of the gain), while the exponentially small $\rho^\mathrm{min}_{N,1}$ in the numerator ensures that the numerator remains close to the disorderless value.

Re-examining $\rho^\mathrm{min}_{N,1}$, we realize that we can reverse the logic that lead to Eq.~\eqref{eq:expansionPBC} and rewrite the series into the inverse of a sum of two matrices
\begin{align}
   \rho^\mathrm{min}_{N,1}
      & = \left\lvert\sum_{n=1}^\infty (-1)^n \big(\big[\chi^{(0)}\big]^{n+1}\big)_{N,1}
          \gamma_\mathrm{eff} w^n \right\rvert
      = \left\lvert
          \sum_{n=0}^\infty (-1)^n \big(\big[\chi^{(0)}\big]^{n+1}\big)_{N,1}
          \gamma_\mathrm{eff} w^n
          - \chi^{(0)}_{N,1}
          \right\rvert \notag \\
      & = \left\lvert
          (M+\gamma_\mathrm{eff} w \mathbb{1})^{-1}
          - \chi^{(0)}
          \right\rvert_{N,1}
      \equiv \left\lvert
          \chi^{(w)}_{N,1}
          - \chi^{(0)}_{N,1}
          \right\rvert \label{eq:rhoMinChi}
\end{align}
with $\bm\chi^{(w)}\equiv (M+\gamma_\mathrm{eff} w \mathbb{1})^{-1}$. In fact, $\rho^\mathrm{min}_{N,1}$ is the difference between a susceptibility matrix $\bm\chi^{(w)}$ where we have subtracted the maximally possible disorder $\gamma_\mathrm{eff} w \mathbb{1}$ from the diagonal and the susceptibility matrix without disorder $\bm\chi^{(0)}$. Since both expressions do not contain random variables any more, we have exact expressions available~\cite{Wanjura2020}. In particular, we have in the topologically non-trivial regime
\begin{align}
   \chi^{(w)}_{N,1} = - \frac{1}{\mu_-} - \varepsilon_{(N-1)}^{(0)} \label{eq:exactIdentity1}
\end{align}
in which the index $(0)$ denotes the value of $\varepsilon_{(N-1)}$ without disorder, which is exponentially small for $\nu\neq0$,
and analogously
\begin{align}
   \chi^{(w)}_{N,1} = - \frac{1}{\mu_-} - \varepsilon_{(N-1)}^{(w)} \label{eq:exactIdentity2}
\end{align}
with disorder of strength $w$.
Overall, we find
\begin{align}
   \rho^\mathrm{min}_{N,1} & = \left\lvert\varepsilon_{(N-1)}^{(0)} - \varepsilon_{(N-1)}^{(w)}\right\rvert. \label{eq:rhoMinTop}
\end{align}
Since both $\varepsilon_{(N-1)}^{(0)}$ and $\varepsilon_{(N-1)}^{(w)}$ are exponentially attenuated with system size, so is $\rho_{N,1}^\mathrm{min}$.

We proceed analogously for $\rho_{N,1}^\mathrm{max}$, although, we first make another estimate to show that $\rho^\mathrm{max}_{N,1}$ is bounded by a function attenuated exponentially with system size from below
\begin{align}
   \rho^\mathrm{max}_{N,1}
      & = \sum_{n=1}^\infty \big(\big\lvert\chi^{(0)}\big\rvert^{n+1}\big)_{N,1}
         \gamma_\mathrm{eff} w^n
      = \sum_{n=0}^\infty \big(\big\lvert\chi^{(0)}\big\rvert^{n+1}\big)_{N,1}
         \gamma_\mathrm{eff} w^n
         - \big\lvert\chi^{(0)}\big\rvert_{N,1}
         \notag \\
      & = \left\lvert (\big\lvert\chi^{(0)}\big\rvert^{-1} - \gamma_\mathrm{eff} w \mathbb{1})^{-1}
      - \big\lvert\chi^{(0)}\big\rvert
      \right\rvert_{N,1} \notag \\
      & \leq \max_\phi \left\lvert \lvert \lvert M\rvert - e^{\i\phi} \, \gamma_\mathrm{eff} w \mathbb{1}\rvert^{-1}
      - \big\lvert\chi^{(0)}\big\rvert
      \right\rvert_{N,1}
      \equiv \left\lvert \lvert\tilde\chi^{(-w)}\rvert_{N,1} - \lvert\chi^{(0)}\rvert_{N,1} \right\rvert. \label{eq:rhoMaxChi}
\end{align}
with $\tilde{\bm\chi}^{(-w)}\equiv \max_\phi \lvert \lvert M\rvert - e^{\i\phi}\gamma_\mathrm{eff} w \mathbb{1}\rvert^{-1}$ (this estimate can be used since the system is translational invariant; $\phi$ is chosen to maximize $(\lvert \lvert M\rvert- e^{\i\phi} \, \gamma_\mathrm{eff} w \mathbb{1}\rvert^{-1})_{N,1}$).

If the matrix $\max_\phi\lvert \lvert M\rvert - e^{\i\phi}\gamma_\mathrm{eff} w \mathbb{1}\rvert$ still is topologically non-trivial, we can, in principle, calculate exact expressions using the results from~\cite{Wanjura2020} inserting the new parameters from $\max_\phi\lvert \lvert M\rvert - e^{\i\phi}\gamma_\mathrm{eff} w \mathbb{1}\rvert$
\begin{align*}
   \tilde\chi^{(-w)}_{N,1} = \max_\phi\lvert \lvert M\rvert - e^{\i\phi} \gamma_\mathrm{eff} w \mathbb{1}\rvert^{-1}_{N,1} & = -\frac{1}{\mu_-} - \tilde\varepsilon_{(N-1)}^{(-w)}
   \label{eq:exactIdentity3}
\end{align*}
in which $\tilde\varepsilon_{(N-1)}^{(-w)}$ is again exponentially suppressed with system size.
Therefore, the overall expression for the bound decays exponentially with system size
\begin{align}
   \rho^\mathrm{max}_{N,1}
   \geq \left\lvert
      \tilde\varepsilon_{N,1}^{(-w)} - \varepsilon_{N,1}^{(0)}
   \right\rvert.
\end{align}
This concludes our argument that the bounds preserve the exponential scaling of the gain with system size for \emph{any} realization of the disorder.

In principle, we can extend this argument of the exponential scaling to the other matrix elements, \eqref{eq:boundsAllElements}, such that we can conclude that the general appearance, i.e. the exponential growth of $\lvert\chi_{j,\ell}\rvert$ with output site $j$ along the chain, is preserved under disorder, see for instance Fig.~\ref{fig:PBCExpIntuitive}~(c) in the main text.


%
%
%
%
%

\end{document}